\def\rn{\noindent\parshape 2 0truecm 8.5truecm 0.3truecm 8.2truecm}
\def\rn{}
\def\nn#1 #2{#2. #1}				
\def\nnn#1 #2 #3{#2. #3. #1}			
\def\nnnn#1 #2 #3 #4{#2. #3. #4 #1}		
\def\nnnnn#1 #2 #3 #4 #5{#2. #3. #4 #5. #1}	
\def\dualand{ and\hbox{ }}				
\def\multiand{, and\hbox{ }}				
\def\rf#1;#2;#3;#4;#5 {{\frenchspacing\par\rn#1, #3 {\bf #4}, #5 (#2). \par}}
\def\rg#1;#2;#3;#4;#5;#6 {{\frenchspacing\par\rn#1, #3 {\bf #4}, #5 (#2). \par}}
\def\rfbook#1;#2;#3;#4;#5 {{\frenchspacing\par\rn#1, {\it #3} (#5, #4, #2).\par}}
\def\rfprep#1;#2;#3 {{\par\frenchspacing\rn#1, #3 (#2).\par}}

\def\preskip {\vskip-0.0cm}
\def\postskip{\vskip+0.1cm}

\def\beq#1{\begin{equation}\label{#1}}
\def\eeq{\end{equation}}
\def\beqa#1{\begin{eqnarray}\label{#1}}
\def\eeqa{\end{eqnarray}}
\def\eq#1{equation~(\ref{#1})~}
\def\Eq#1{Equation~(\ref{#1})~}

\def\Cbf {{\bf C}}

\def\zbf {{\bf z}}
\def\ybf {{\bf y}}
\def\xbf {{\bf x}}
\def\x{{\bf x}}

\def\albf{{\rm a}}
\def\zbft{{\bf z}^T}

\def\ybft{{\bf y}^T}

\def\albfHat{\widehat{{\rm a}}} 
\def\da{\Delta\albf}

\def\NN{\rm \bf N}
\def\SCMB{\rm \bf S}
\def\W{\rm \bf W}
\def\xw{{\xbf}^w}
\def\yw{{\ybf}^w}

\def\tr{\hbox{tr}\,}

\def\arcdeg{^\circ}

\def\etal{{\frenchspacing\it et al.}}
\def\ie{{\frenchspacing\it i.e.}}
\def\eg{{\frenchspacing\it e.g.}}

\def\expec#1{\langle#1\rangle}
\def\l{\ell}

\def\spose#1{\hbox to 0pt{#1\hss}}
\def\simlt{\mathrel{\spose{\lower 3pt\hbox{$\mathchar"218$}}
    \raise 2.0pt\hbox{$\mathchar"13C$}}}
\def\simgt{\mathrel{\spose{\lower 3pt\hbox{$\mathchar"218$}}
    \raise 2.0pt\hbox{$\mathchar"13E$}}}
\def\simpropto{\mathrel{\spose{\lower 3pt\hbox{$\mathchar"218$}}
    \raise 2.0pt\hbox{$\propto$}}}

\def\fig#1{Figure~\ref{#1}}
\def\Fig#1{Figure~\ref{#1}}

\def\sec#1{Section~\ref{#1}}

\def\draft{
}

\documentstyle[prd,aps,epsf]{revtex}
\draft
\begin{document}
\twocolumn[\hsize\textwidth\columnwidth\hsize\csname@twocolumnfalse\endcsname



\title{How Accurately can Suborbital Experiments Measure the CMB?}


\author{Ang\'elica de Oliveira-Costa$^1$, Max Tegmark$^1$, Mark J. Devlin$^1$, Lyman Page$^2$,  Amber D. Miller$^3$, \\
        C.~Barth Netterfield$^4$, Yongzhong Xu$^5$}

\address{$^1$Dept. of Physics, University of Pennsylvania, Philadelphia, PA 19104;angelica@hep.upenn.edu}
\address{$^2$Dept. of Physics, Princeton University, Princeton, NJ 08544} 
\address{$^3$Dept. of Physics, Columbia University, New York, NY 10027}
\address{$^4$Dept. of Physics and Astronomy, University of Toronto, Toronto, ON M5S1A7, Canada}
\address{$^5$Los Alamos Natl. Lab., P.O.Box 1663, Los Alamos, MN 87545}

\date{\today. Submitted to PRD.}

\date{\today}

\maketitle

\begin{abstract}
Great efforts are currently being channeled into ground- and balloon-based 
CMB experiments, mainly to explore anisotropy on small angular scales and polarization.
To optimize instrumental design and assess experimental prospects, it is important
to understand in detail the atmosphere-related systematic errors that 
limit the science achievable with new suborbital instruments.
For this purpose, we spatially compare the 648 square degree ground- and 
balloon-based QMASK map with the atmosphere-free WMAP map, finding beautiful 
agreement on all angular scales where both are sensitive.
This is a reassuring quantitative assessment of the power of the 
state-of-the-art FFT- and matrix-based mapmaking techniques that have been 
used for QMASK and virtually all subsequent experiments. 
\end{abstract}



] 


\section{INTRODUCTION}

Since the discovery of Cosmic Microwave Background (CMB) temperature anisotropy 
by COBE/DMR \cite{smoot92},
a variety of {\it ground-based} and {\it balloon-borne} (or {\it suborbital}) experiments have 
detected and characterized the fluctuations in the CMB on different angular scales. 
These suborbital observing locations were chosen for being less costly than space-based 
alternatives, but at a price: none of these experiments could enjoy the level 
of systematic control and calibration accuracy that has been achieved in space 
missions such as COBE and WMAP \cite{bennett03a}. To deal with the contamination 
from both atmospheric emission and various other offsets and modulations related 
to the warmer and less stable suborbital environment, a powerful set of data 
analysis tools were developed. These come in many names and guises 
(see, \eg, 
\cite{mapmaking,guide,Netterfield97,hacme,qmap1,qmap2,qmap3,Borrill99,deBernardis00,Hanany00,Stompor02,Wandelt01,Halverson02,Hivon02,Myers03,Hinshaw03}), 
but all do roughly the same thing. First the raw time-ordered data is cleaned 
both in real space (removing segments of low-quality data) and in Fourier space 
(using notch filters to remove, \eg, scan-synchronous offsets and balloon 
pendulation lines) and approximately prewhitened with FFT techniques to better 
treat correlated noise and detector response.
Then the time-ordered data is compressed into a pixelized sky map by inverting 
a large matrix, either by brute force, iteratively or using some  
approximation, and the corresponding noise covariance matrix is calculated or 
estimated. Numerous tricks for numerically accelerating this second step are 
available using fast Fourier transforms, circulant matrices or Toeplitz matrices.

Although they are mathematically elegant, there have been few precision tests 
of how well these methods work in practice for suborbital experiments. The 
numerical methods on which they are based only guarantee that they can accurately 
deal with precisely those problems that they were designed to deal with, so 
it is crucial 

\begin{figure}[tb]
\preskip
\centerline{\epsfxsize=8.5cm\epsffile{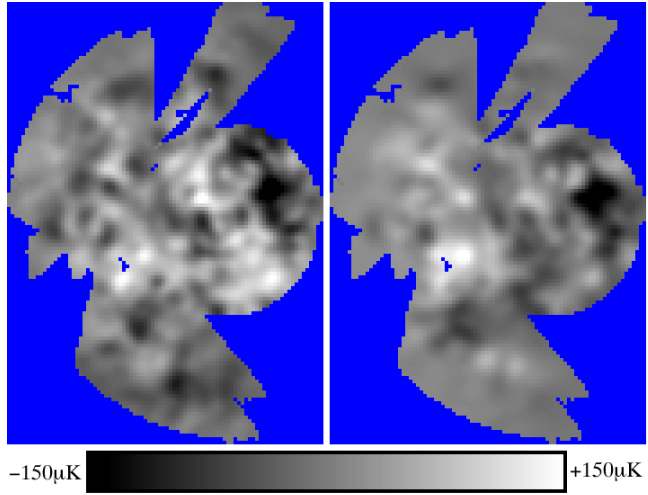}}
\medskip
\caption{\label{visualcomparison}\footnotesize%
	Visual comparison of the QMASK (left) and WMAP (right) maps,
	both Wiener filtered with the same weighting 
        (determined by the QMASK noise covariance matrix).
	The slightly larger 
	fluctuation level
	seen in the left
	image gives a visual indication of the effect of QMASK noise.
	 }
\end{figure} 
\medskip

\noindent
to perform end-to-end tests that are able to detect unforeseen problems as well. 
By far the most powerful test (in the statistical sense of  being more likely to 
discover systematic errors) involves a direct comparison 
of the sky maps from experiments that overlap in both spatial and angular 
coverage, rather than merely a comparison of the measured angular power spectra.
Some of the best testimony to the quality of CMB maps comes from the success of 
such map comparisons in the past --- 
	between FIRS and DMR \cite{Ganga93},
	Tenerife and DMR \cite{Lineweaver95},
	FIRAS and DMR \cite{hinshaw96}, 
	MSAM and SK (Saskatoon) \cite{Knox98},
        QMAP and SK \cite{costa98},
	QMASK and DMR \cite{qmask}, 
        PIQUE and SK \cite{costa02},
        POLAR and DMR \cite{polarcosta03},
        DMR and WMAP \cite{bennett03a},
        Maxima and WMAP \cite{abroe04}, 
        BOOMERanG and WMAP \cite{deBernardis03},
        Tenerife and WMAP \cite{costa03},
	two years of Python data \cite{Ruhl95},
        four years of DMR data \cite{bennett96}, 
	three years of SK data \cite{saskmap},
	two flights of  MSAM \cite{Inman97}, 
and different channels of 
	QMAP \cite{qmap1,qmap2,qmap3},
	BOOMERanG \cite{deBernardis00},
	MAXIMA \cite{Hanany00}, and
	WMAP \cite{Hinshaw03}.
It has now become possible to perform such tests with unprecedented sensitivity,
using the spectacularly clean and well-tested WMAP data \cite{bennett03a} as 
the answer sheet against which to compare other maps. So far, this has only 
been done carefully (beyond simple visual comparisons) for MAXIMA, with 
encouraging results \cite{abroe04}.
This work showed that the WMAP-MAXIMA cross power spectrum was consistent with the 
power spectra measured from the two experiments separately.

The purpose of this letter is to carry out such a test for the QMASK data, which
was the largest publicly available degree-scale CMB map prior to the WMAP release,
and covers more than six times the area of the MAXIMA-I map. Not only was 
this map created using the above-mentioned Fourier and matrix techniques,
but it combines measurements from both the ground-based SK experiment 
\cite{Netterfield97} and two balloon-based data sets from the QMAP experiment 
\cite{qmap1,qmap2,qmap3}, therefore being potentially vulnerable to almost any 
systematic error one could imagine. 

Although the two maps have different angular resolutions, their comparison is 
straightforward since the WMAP data is effectively noiseless (compared to the 
QMASK map) and can easily be smoothed to the QMASK angular resolution.  
Below we compare these two data sets using four different statistical tests, then 
present our conclusions in \sec{ConclusionsSec}.

\section{METHODS \& RESULTS }\label{METHODSec}

\subsection{Data sets used in the analysis}

The QMASK map was constructed from the combination of the SK and QMAP 
data sets \cite{qmask}. It consists of 6590 pixels arranged on a simple square 
grid in gnomonic equal area projection, and its angular resolution corresponds 
to a Gaussian beam with FWHM $0.68\arcdeg$.
These pixels are conveniently arranged in a 6590-dimensional vector $\x$ whose
noise covariance matrix $\NN_{\rm QMASK}\equiv\expec{\x\x^t}$ is known and available 
online \cite{qmask}. We compare it with both the WMAP Q-band map \cite{bennett03a} 
and a foreground-cleaned CMB map made from a combination of the five WMAP CMB maps 
\cite{tegmark03b}, hereafter TOH03. Before performing any analysis, the WMAP-Q and 
TOH03 maps were smoothed to have the same net beam function as QMASK, and data that 
do not overlap the QMASK observing region were discarded.  
 
\subsection{Visual Comparison}

We begin with a simple visual comparison. Comparing plots of the raw maps of QMASK 
($\xbf$) and WMAP ($\ybf$) is not particularly useful, since QMASK is a noisy data 
set. Instead, we compare Wiener-filtered versions of both maps:
\beq{wienermaps}
    \xw=\W\xbf,\quad \yw=\W\ybf,             \nonumber 
\eeq
where 
\beq{Weq}
    \W=\SCMB [\SCMB + \NN_{\rm QMASK} + \NN_{\rm WMAP} ]^{-1}, \nonumber 
\eeq
and $\SCMB$ is the contribution to the map covariance matrix from CMB signal.
The WMAP noise is so much smaller than that of QMASK that 
its exact value is irrelevant for our applications, so we make
the approximation  
$\NN_{\rm WMAP} \approx 0$, \ie, 
\beq{Weq2}
   \W \approx \SCMB [\SCMB + \NN_{\rm QMASK}]^{-1}. \nonumber 
\eeq
This linear filtering maximizes the signal-to-noise ratio in the QMASK
map by effectively smoothing the map and downweighting noisy pixels and
modes. \Eq{wienermaps} shows that to ensure a fair comparison, we Wiener-filter
the WMAP map using the exact same $\W$-matrix as for the QMASK map.
The Wiener filtered QMASK map $\xw$ (left) and WMAP TOH03 map $\yw$ (right) are shown
in \Fig{visualcomparison}, and are seen to look encouragingly similar. Even the 
``anomalous'' cold-spot around (RA,DEC)=($3^{h}20^{m}$, $84\arcdeg 55^{'}$)
discussed in the QMASK analysis \cite{qmask} 
(also seen by SK \cite{wollack93} but not by DMR \cite{qmask}) 
is clearly visible in 
the WMAP data. 
The results are found to depend only weakly on the detailed choice of $\SCMB$. We wish to keep 
our map comparisons as theory-free as possible, and therefore choose $\SCMB$ corresponding 
to a simple scale-invariant spectrum with $\delta T_\ell=30\mu K$ here and below. 
 
\subsection{Cross-correlation results}

Next, we compare the QMASK and WMAP maps by cross-correlating them.
The measured cross-correlation coefficient $\albfHat$ and its variance $\da$ are given by 
\cite{saskforeg}
\beq{alpha}
   \albfHat = {\ybft \Cbf^{-1} \xbf\over\ybft \Cbf^{-1} \ybf},
\eeq
\beq{varalpha}
   \da^2 \equiv \expec{\albfHat^2} - 
                \expec{\albfHat}^2 =
         \left[ \ybft \Cbf^{-1} \ybf \right]^{-1},
\eeq
where the covariance matrix $\Cbf$ is defined as 
\beq{covarmatrix}
	\Cbf \equiv  \NN + \SCMB =
        \NN_{\rm QMASK} +                  \SCMB.
\eeq

As a first reality check, we cross-correlate the full QMASK data with the TOH03 map, 
obtaining $\albfHat=0.921\pm 0.061$. This provides an encouraging validation of the 
QMASK observations, showing a detection of CMB signal at about the $15\sigma$  
level. This also confirms and further strengthens the results of \cite{qmask}, who 
cross-correlated QMASK and DMR data \cite{smoot92,bennett96} and concluded that 
the CMB signal was consistent between the two data sets.
It also provides a completely independent test of the accuracy with which the 
QMAP and SK experiments were calibrated, showing that they got things 
right to well within the quoted 10\% calibration uncertainty \cite{qmask}.
 
Using only the sky area above $|b|>20^{\circ}$ (a Galactic cut), we detect 
the CMB signal at the $14.7\sigma$ level 
($\albfHat=0.938\pm 0.064$), and 
for a $|b|>30^{\circ}$ Galactic cut the detection is around the $9\sigma$ level
($\albfHat=0.950\pm 0.110$). The similarity between the 20$^{\circ}$ and 30$^{\circ}$ 
cut results also confirms the conclusion of \cite{qmapforeg} that QMASK foregrounds 
are not dominant at high Galactic latitudes ($b>$20$^{\circ}$).

\subsection{The $\chi^2$-test} 

A simple and robust way to test for discrepancies between the QMASK and WMAP maps 
is to calculate $\chi^2$ defined by
\beq{chinoise}
    \chi^2 \equiv \zbft  \NN^{-1} \zbf  \approx 
    \albfHat^{-2} \zbft \NN^{-1}_{\rm QMASK}  \zbf , 
\eeq
where $\zbf  = \albfHat\xbf  - \ybf$ is the difference of the two maps after 
adjusting the QMASK calibration to the best-fit value found in the previous section.
In the absence of systematic or calibration errors, the corresponding difference 
map should contain pure noise, so we should have 
	$\expec{\zbf}$=0 and $\chi^2/N_{pix} \approx 1$, 
where 
$N_{pix}=6590$ is the number of pixels in the QMASK map vector $\xbf$. This is 
precisely what we find: we obtain $\chi^2/N_{pix}\approx 1.00$, \ie, the difference 
map $\zbf$ is consistent with pure noise. This means that the strong signal seen 
in \fig{visualcomparison} is common to both maps and due to a true sky signal, 
with no evidence for significant systematic errors in QMASK.

\subsection{The Null-buster test}

How statistically significant is our detection of signal in the maps?
Consider the null hypothesis that a map $\zbf$ contains merely noise,
\ie, $\expec{\zbf \zbft}=\NN$. Given the alternative hypothesis 
$\expec{\zbf \zbft}=\NN+\SCMB$, one can show that the most powerful
``null-buster'' test for ruling out the null hypothesis is using 
the statistic \cite{comparing,qmask}
\beq{NullbusterEq}
    \nu \equiv {\zbft \NN^{-1} \SCMB \NN^{-1} \zbf - \tr\NN^{-1} \SCMB
    \over
    \left[2\,\tr\left\{\NN^{-1} \SCMB \NN^{-1} \SCMB \right\}\right]^{1/2}},
\eeq
which can be interpreted as the number of ``sigmas'' at which the
null noise-only hypothesis is ruled out. 
Whenever we have reason to suspect systematic errors of a certain
form are producing a signal $\propto\SCMB$, the null-buster test is  
more sensitive to these systematic errors than the 

\begin{figure}[tb]
\preskip
\centerline{\epsfxsize=8.5cm\epsffile{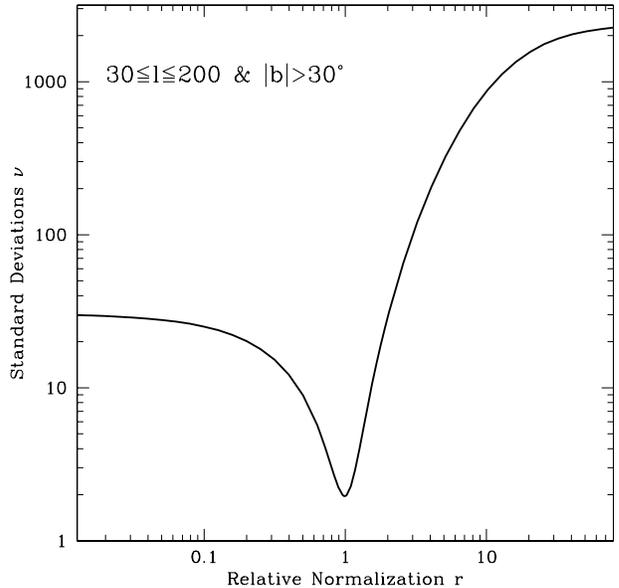}}
\postskip
\caption{\label{nullbuster}\footnotesize%
	Null-buster comparison of QMASK (left) and WMAP-Q (right). 
	This figure shows the number of standard deviations (``sigmas'') 
	at which the difference map $\zbf = \xbf  - {\rm r} \ybf$ is inconsistent 
	with pure noise. }
\end{figure} 
\bigskip

\noindent
$\chi^2$-test described in the previous subsection, which is a general-purpose tool.

The presence of foreground contamination (see \cite{foregpars}
for a review) has been quantified independently for both the 
SK \cite{saskforeg} and QMAP \cite{qmapforeg} experiments, by 
cross-correlating them with various foreground templates.
The contamination level was found to be small enough to have 
negligible impact on the measured QMASK power spectrum \cite{qmaskpow},
but as we will see below, it is nonetheless large enough to affect 
the more sensitive test we are performing here. 
The main QMASK foregrounds are synchrotron radiation, free-free emission and
perhaps spinning dust emission, all of which become more important (relative 
to the CMB) at lower galactic latitudes and on larger angular scales (low $\l$).
To focus on the CMB signal, Figure 2 therefore shows the nullbuster test results 
limited to $|b|>30^\circ$ and angular scales $\l\ge 30$. The latter is 
achieved by summing only over $\l=30,...,200$ when computing $\SCMB$ in 
\eq{NullbusterEq} (beam issues limit QMASK to $\l\simlt 200$ \cite{qmaskpow}). 
This also corresponds to the angular range in which QMASK carries the bulk 
of its information, since limited sky coverage and angular resolution
limit the information at $\ell\ll 30$ and $\l\gg 200$, respectively.

\Fig{nullbuster} shows that whereas QMASK (left; $r=0$) and WMAP-Q (right; 
$r=\infty$) detect sky signal at the level of about $30\sigma$ and $2300\sigma$, 
respectively, the difference map (corresponding to $r$=1) is consistent with 
pure noise, and the null hypothesis of no systematic errors fails at less than 
$2\sigma$ level.\footnote{For this test, we made the approximation that the 
WMAP noise was uncorrelated between pixels with a standard deviation 
$\approx 16\mu$K 
estimated by subtracting the two Q-band WMAP channels from each other.
}
The fact that the curve has such a narrow minimum at $r\sim 1$ also provides 
a strong confirmation that the relative calibration of QMASK and WMAP is good.

We used the WMAP-Q map for this test because it is closest in frequency
to that of the QMASK map, which is constructed from a combination of Ka- and 
Q-band data. We also repeated the nullbuster test using the foreground-cleaned
TOH03 WMAP map, finding marginally worse agreement as expected, since this 
compares a map that includes foregrounds with an essentially foreground-free map.
As expected, the agreement between the two maps also deteriorated when including 
data at lower Galactic latitudes, where foregrounds become more important.

  
\section{CONCLUSIONS}\label{ConclusionsSec}

In summary, a powerful set of FFT- and matrix-based tools have been developed 
in recent years for producing quality CMB maps from time-ordered observations. 
Although these tools are mathematically elegant, there have been few 
precision tests of how well they work in practice when faced with real-world 
systematic errors from suborbital experiments.
We have performed a precision test of precisely this, comparing the QMASK map 
with the ``answer sheet'' provided by WMAP. The results are very encouraging, 
showing excellent agreement and no statistically significant evidence of 
residual systematic errors.
This suggests that although space-based observations will obviously be required 
for many tasks, \eg, all-sky mapping of CMB polarization, suborbital experiments 
have the potential to continue playing a key role in the precision cosmology era.

\medskip

The authors wish to thank David Spergel for useful comments
and the WMAP team for producing such a superb
data set and making it public via the Legacy Archive for Microwave Background Data 
Analysis (LAMBDA), 
supported by the NASA Office of Space Science.
This work was supported by NASA grant NAG5-11099,
NSF grants AST-0134999 and AST9732960, and fellowships from the David and Lucile
Packard Foundation and the Research Corporation.  


\end{document}